\documentclass[preprint,12pt, a4paper]{elsarticle}

\usepackage[T1]{fontenc}
\usepackage[utf8]{inputenc}
\usepackage{amssymb}
\usepackage{listings}
\usepackage{multirow}
\usepackage{hyperref}
\usepackage{cleveref}

\journal{SoftwareX}

\begin{document}

\newcommand{\q}[1]{``#1''}
\renewcommand{\labelenumii}{\arabic{enumi}.\arabic{enumii}}
\newcommand{\skfp}{\textit{scikit-fingerprints}}

\interfootnotelinepenalty=10000

\begin{frontmatter}



\title{Scikit-fingerprints: easy and efficient computation of molecular fingerprints in Python}


\author{Jakub Adamczyk\corref{cor1}\fnref{label1}}
\ead{jadamczy@agh.edu.pl}

\author{Piotr Ludynia\fnref{label2}}

\cortext[cor1]{Corresponding author}
\address{AGH University of Krakow, Department of Computer Science, Cracow, Poland}

\fntext[label1]{ORCID 0000-0003-4336-4288}
\fntext[label2]{ORCID 0009-0004-0749-9569}

\begin{abstract}
In this work, we present \skfp, a Python package for computation of molecular fingerprints for applications in chemoinformatics. Our library offers an industry-standard scikit-learn interface, allowing intuitive usage and easy integration with machine learning pipelines. It is also highly optimized, featuring parallel computation that enables efficient processing of large molecular datasets. Currently, \skfp~stands as the most feature-rich library in the open source Python ecosystem, offering over 30 molecular fingerprints. Our library simplifies chemoinformatics tasks based on molecular fingerprints, including molecular property prediction and virtual screening. It is also flexible, highly efficient, and fully open source.
\end{abstract}


\end{frontmatter}

\section*{Metadata}
\label{}

\begin{table}[!h]
\resizebox{\textwidth}{!}{
\begin{tabular}{|l|p{6.5cm}|p{6.5cm}|}
\hline
\textbf{Nr.} & \textbf{Code metadata description} & \textbf{Please fill in this column} \\
\hline
C1 & Current code version & 1.6.1 \\
\hline
C2 & Permanent link to code/repository used for this code version & \url{https://github.com/scikit-fingerprints/scikit-fingerprints/tree/SoftwareX_submission_v1.6.1} \\
\hline
C3  & Permanent link to Reproducible Capsule & N/A \\
\hline
C4 & Legal Code License   & MIT \\
\hline
C5 & Code versioning system used & git \\
\hline
C6 & Software code languages, tools, and services used & Python 3.9 or newer, RDKit \\
\hline
C7 & Compilation requirements, operating environments \& dependencies & Linux, Windows, macOS \\
\hline
C8 & If available Link to developer documentation/manual &  \url{https://scikit-fingerprints.readthedocs.io/latest/} \\
\hline
C9 & Support email for questions & jadamczy@agh.edu.pl \\
\hline
\end{tabular}
}
\caption{Code metadata}
\label{codeMetadata} 
\end{table}

\section{Motivation and significance}

Molecules are the basic structures processed in computational chemistry. They are most commonly represented as molecular graphs, which need to be converted into multidimensional vectors for the majority of processing algorithms, most prominently for machine learning (ML) applications. This is typically done with molecular fingerprints, which are feature extraction algorithms that encode structural information about molecules as vectors \cite{molecular_descriptors_in_cheminformatics}. They are widely used in chemoinformatics, e.g. for chemical space diversity measurement \cite{chemical_space_diversity,diversity_picking,molecule_similarity} and visualization \cite{visualization,visualization_2}, clustering \cite{clustering,clustering_2,clustering_3,clustering_and_diversity_picking}, virtual screening \cite{virtual_screening,virtual_screening_2}, molecular property prediction \cite{fingerprints_applications,fingerprints_applications_4,fingerprints_concatenation}, and many more \cite{fingerprints_applications_2,fingerprints_applications_3,fingerprints_applications_5,fingerprints_applications_6,fingerprints_applications_7,fingerprints_applications_8}. These chemoinformatics tasks, which often rely on machine learning methods, are important for many real-life applications, particularly drug design. For properly assessing the performance of predictive models, train-test splitting is crucial, and molecular fingerprints can also be used there \cite{splitting_data,split_maxmin,split_maxmin_2,split_maxmin_3,split_simpd}. The performance of fingerprint-based models remains very competitive, even compared to state-of-the-art graph neural networks (GNNs) \cite{fingerprints_applications_4}. Hybrid molecular property prediction models are also a subject of recent research, combining molecular fingerprints with GNNs \cite{fingerprints_and_GNNs,fingerprints_and_GNNs_2,fingerprints_and_GNNs_3,fingerprints_and_GNNs_4}, transformers \cite{fingerprints_and_transformers,fingerprints_and_transformers_2}, or autoencoders \cite{fingerprints_and_autoencoders}.

The selection of the optimal fingerprint representation for a given application is nontrivial. It typically requires the computation of many different fingerprints \cite{fingerprints_applications_4}, and may also require tuning their hyperparameters \cite{fingerprints_hyperparameters,fingerprints_hyperparameters_2}. Using multiple fingerprints at once often improves results, e.g. via concatenation \cite{fingerprints_concatenation} or data fusion \cite{data_fusion,data_fusion_2}. Processing large molecular datasets necessitates efficient implementations that leverage modern multicore CPUs. Python, the most popular language in chemoinformatics today, includes the scikit-learn library \cite{scikit-learn}, which has become the de facto standard tool for tabular machine learning tasks, and deep learning frameworks like PyTorch \cite{pytorch}. Scikit-learn in particular is renowned for its intuitive and widely adopted API \cite{scikit-learn-api}.

Popular open source tools for computing molecular fingerprints, such as Chemistry Development Kit (CDK) \cite{CDK}, Open Babel \cite{OpenBabel}, and RDKit \cite{RDKit}, are written in Java or C++. None of them are compatible with the scikit-learn API, and their Python wrappers can be cumbersome to work with. They also offer no or very limited support for parallel computation.

Here, we present \skfp, a new Python library for easy and efficient computation of molecular fingerprints. It is fully scikit-learn compatible, enabling easy integration into ML pipelines as a feature extractor for molecular data. It offers optimized parallel computation of fingerprints, enabling the processing of large datasets and experiments with multiple algorithms. We implemented over 30 different fingerprints, making it the most feature-rich library in the open source Python ecosystem for molecular fingerprinting. Those include those based only on molecular graph topology (2D), as well as those utilizing graph conformational structure (3D, spatial). It is fully open source, publicly available on PyPI \cite{python_pypi} and on GitHub at \href{https://github.com/scikit-fingerprints/scikit-fingerprints}{https://github.com/scikit-fingerprints/scikit-fingerprints}.

\section{Software description}

\subsection{Software architecture}
\begin{figure}
    \centering
    \includegraphics[width=\textwidth]{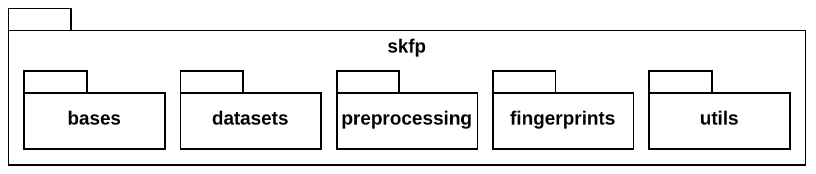}
    \caption{Package diagram of \skfp.}
    \label{fig:package-diagram}
\end{figure}

\skfp~is a Python package for computing molecular fingerprints, designed for chemoinformatics and ML workflows. Its interface is fully compatible with scikit-learn API \cite{scikit-learn-api}, ensured by proper inheritance from scikit-learn base classes and comprehensive tests.

The package structure is shown in Figure \ref{fig:package-diagram}. All functionality is contained in the \texttt{skfp} package, allowing easy imports. The base classes are in \texttt{skfp.bases} package, and they can be used to extend the functionality with new or customized fingerprints. \texttt{skfp.datasets} has functions to load popular datasets for easy benchmarking. \texttt{skfp.preprocessing} contains classes for preprocessing molecules before computing fingerprints, as described in Section \ref{subsection_preprocessing}. Fingerprints are represented as classes in package \texttt{skfp.fingerprints}. Lastly, \texttt{skfp.utils} contains additional utilities, such as input type validators.

\subsection{Software functionalities}

User-facing functionalities can be divided into preprocessing and fingerprint calculation. It also supports loading popular datasets. In addition, in contrast to existing software, we support efficient parallelism and implement multiple measures for ensuring high code quality and security.

\subsubsection{Preprocessing}
\label{subsection_preprocessing}

Fingerprints take RDKit \texttt{Mol} objects as input to the \texttt{.transform()} method. However, for convenience, all 2D-based fingerprints also take the SMILES input, converting them internally. If done multiple times, this entails a small performance penalty, so \skfp~offers \texttt{MolFromSmiles} and \texttt{MolToSmiles} classes for easier conversions.

SMILES representation for a molecule is not unique, and there are various non-standard extensions to this format \cite{SMILES_extensions,SMILES_extensions_2, SMILES_extensions_3}. In particular, incorrect or very unlikely molecules can be written in SMILES form. For example, string \q{\texttt{[H]=[H]}} is a syntactically correct SMILES, but is not a chemically valid molecule. \texttt{MolFromSmiles} by design performs only basic sanitization checks, to enable reading arbitrary data. For expanded checks, we implement the \texttt{MolStandardizer} class. Since there is no one-size-fits-all solution for molecular standardization, we use the most widely used standardization steps, recommended by RDKit \cite{RDKit_sanitization}. This helps ensure high data quality at the beginning of the pipeline.

All fingerprints utilizing conformational (3D, spatial) information require \texttt{Mol} input, with conformers calculated using RDKit, with \texttt{conf\_id} property set. Conformer generation can be troublesome, with multiple different algorithms and settings available. \texttt{ConformerGenerator} class in \skfp~greatly simplifies this process, offering reasonable defaults. It attempts to maximize efficiency for easy molecules and minimize the chance of failure for complex compounds, based on the ETKDGv3 algorithm \cite{ETKDGv3}, known to give excellent results \cite{ETKDG_v3_quality}.

\subsubsection{Fingerprints calculation}

\begin{figure}
    \centering
    \includegraphics[width=\textwidth]{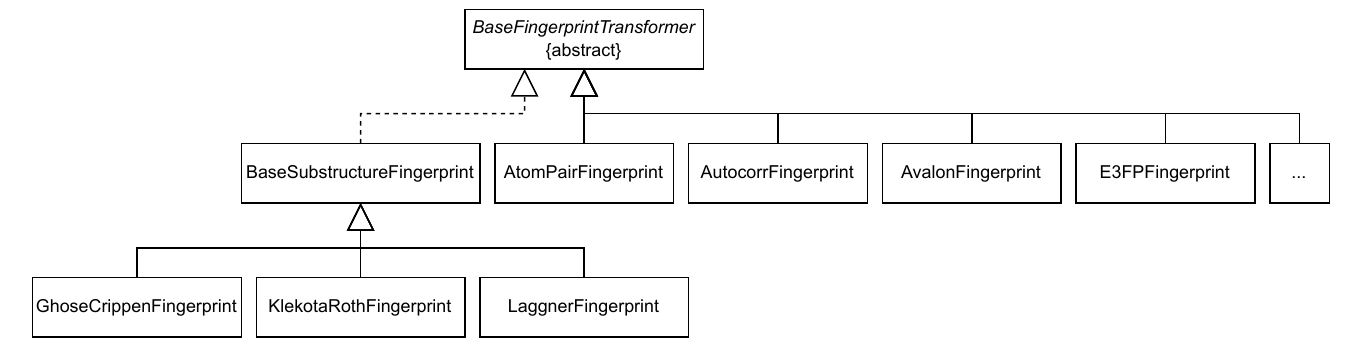}
    \caption{Class diagram for fingerprint classes. Some classes omitted for readability.}
    \label{fig:class-diagram}
\end{figure}

Different molecular fingerprints are represented as classes, all inheriting from \texttt{BaseFingerprintTransformer}, and further from \texttt{BaseSubstructureFingerprint} for substructure fingerprints such as Klekota-Roth \cite{fp_klekota_roth} (see Figure \ref{fig:class-diagram}). They are used as stateless transformers in scikit-learn and used mainly via the \texttt{.transform()} method. It takes a list of SMILES strings or RDKit \texttt{Mol} objects, and outputs a dense NumPy array \cite{numpy} or a sparse SciPy array in CSR format \cite{scipy}. Various options, such as vector length for hashed fingerprints (e.g. ECFP \cite{fp_ecfp}), binary/count variant, dense/sparse output etc. are specified by constructor parameters. This ensures full composability with scikit-learn constructs like pipelines and feature unions.

We implement more than 30 different fingerprints of various types, e.g. circular ECFP \cite{fp_ecfp} and SECFP \cite{fp_mhfp_secfp}, path-based Atom Pair \cite{fp_atom_pair} and Topological Torsion \cite{fp_topological_torsion}, substructure-based MACCS \cite{fp_maccs} and Klekota-Roth \cite{fp_klekota_roth}, physicochemical descriptors such as EState \cite{fp_estate} and Mordred \cite{fp_mordred}, and more. We used efficient RDKit subroutines, written in C++, e.g. for matching SMARTS patterns. A complete list of implemented fingerprints is available in \skfp~\href{https://scikit-fingerprints.github.io/scikit-fingerprints/modules/fingerprints.html}{online documentation}.

\subsubsection{Parallelism}

Since molecules can be processed independently when computing fingerprints, the task is embarrassingly parallel \cite{embarrassingly_parallel}. This means that we can efficiently utilize all available CPU cores. To minimize inter-process communication, by default, input molecules are split into as many chunks as there are cores available, and processed in parallel by Python workers. We utilize Joblib \cite{python_joblib}, with the Loky executor, which uses memory mapping to efficiently pass the resulting arrays between processes. Furthermore, by using sparse arrays and smaller chunk sizes, users can minimize memory utilization for large datasets and fingerprints that yield long output vectors \cite{fingerprints_hyperparameters_2}.

Furthermore, we support distributed computing with Dask \cite{python_dask}, used as a Joblib executor. This way, \skfp~can take advantage of large high-performance computing (HPC) clusters. Connecting to the Dask cluster only requires setting a single parameter in the Joblib configuration \cite{python_joblib_dask_integration}.

\subsubsection{Datasets loading}
\label{subsection_datasets}

Fingerprints are often used in the context of molecular property prediction on standardized benchmarks. In particular, they constitute strong baselines, often outperforming complex graph neural networks (GNNs) \cite{fingerprints_applications,fingerprints_applications_3,fingerprints_applications_4}. Therefore, their easy usage is important for a fair evaluation of advances in graph classification.

We utilized HuggingFace Hub \cite{huggingface,huggingface_hub} to host datasets. It offers easy downloading, caching, and loading datasets, with automated compression to Parquet format. Currently, the most widely used MoleculeNet \cite{MoleculeNet} benchmark has been integrated, and additional datasets can be easily added with the unified interface. Users can load data sets as in scikit-learn. For example, loading the MoleculeNet BBBP dataset uses the function \texttt{load\_bbbp()}.

\subsubsection{Code quality and CI/CD}

We ensure high code quality and security with multiple measures. The code is versioned using Git and GitHub. New features have to be submitted through Pull Requests and undergo code review. We use pre-commit hooks \cite{pre_commit} to verify code quality before each commit:
\begin{itemize}
    \item \texttt{bandit} \cite{python_bandit}, \texttt{safety} \cite{python_safety} - security analysis and dependency vulnerability scanning, following security recommendations \cite{python_bandit_recommendation,python_vulnerability_analaysis_recommendation}
    \item \texttt{black} \cite{python_black}, \texttt{flake8} \cite{python_flake8}, \texttt{isort} \cite{python_isort}, \texttt{pyupgrade} \cite{python_pyupgrade} - code style, following reproducibility and readability guidelines \cite{python_code_linter_recommendation}
    \item \texttt{mypy} \cite{python_mypy} - type checking; our entire code is statically typed, following security recommendations \cite{python_static_checkers_recommendation,python_static_checkers_recommendation_2}
    \item \texttt{xenon} \cite{python_xenon} - cyclomatic complexity
\end{itemize}

We implemented a comprehensive suite of 196 integration and unit tests. They use the pytest framework \cite{python_pytest}, and are run automatically on GitHub Runners as a part of the CI/CD process. Passing all tests is required to merge the code into the master branch. We run tests on a full matrix of operating systems (Linux, Windows, macOS) and Python versions (from 3.9 to 3.12), ensuring proper execution in different environments.

Any changes to the documentation are automatically deployed to the GitHub Pages. New package versions are deployed to PyPI by using GitHub Releases, with new changes description. Internally, this uses a GitHub Actions workflow and creates a Git tag on the commit used in the given release. \skfp~can be installed with pip by running \texttt{pip install scikit-fingerprints}.

\section{Illustrative examples}

\subsection{Parallel computation}

Since computing molecular fingerprints is an embarrassingly parallel task, it can very effectively utilize modern multicore CPU architectures, e.g. for large databases in molecular property prediction or virtual screening. To illustrate the capability of \skfp~in this regard, we calculate fingerprints for the popular HIV dataset from the MoleculeNet benchmark \cite{MoleculeNet}. It contains a wide variety of molecules from medicinal chemistry, including organometallics, small and large molecules, some atoms with very high numbers of bonds, etc. For this experiment, we limit the data to 10 thousand molecules, due to the high computational time required to run the benchmark multiple times for many data sizes and fingerprints. The code is available in the GitHub repository, in \texttt{benchmarking} directory.

\begin{figure}
    \centering
    \includegraphics[width=\textwidth]{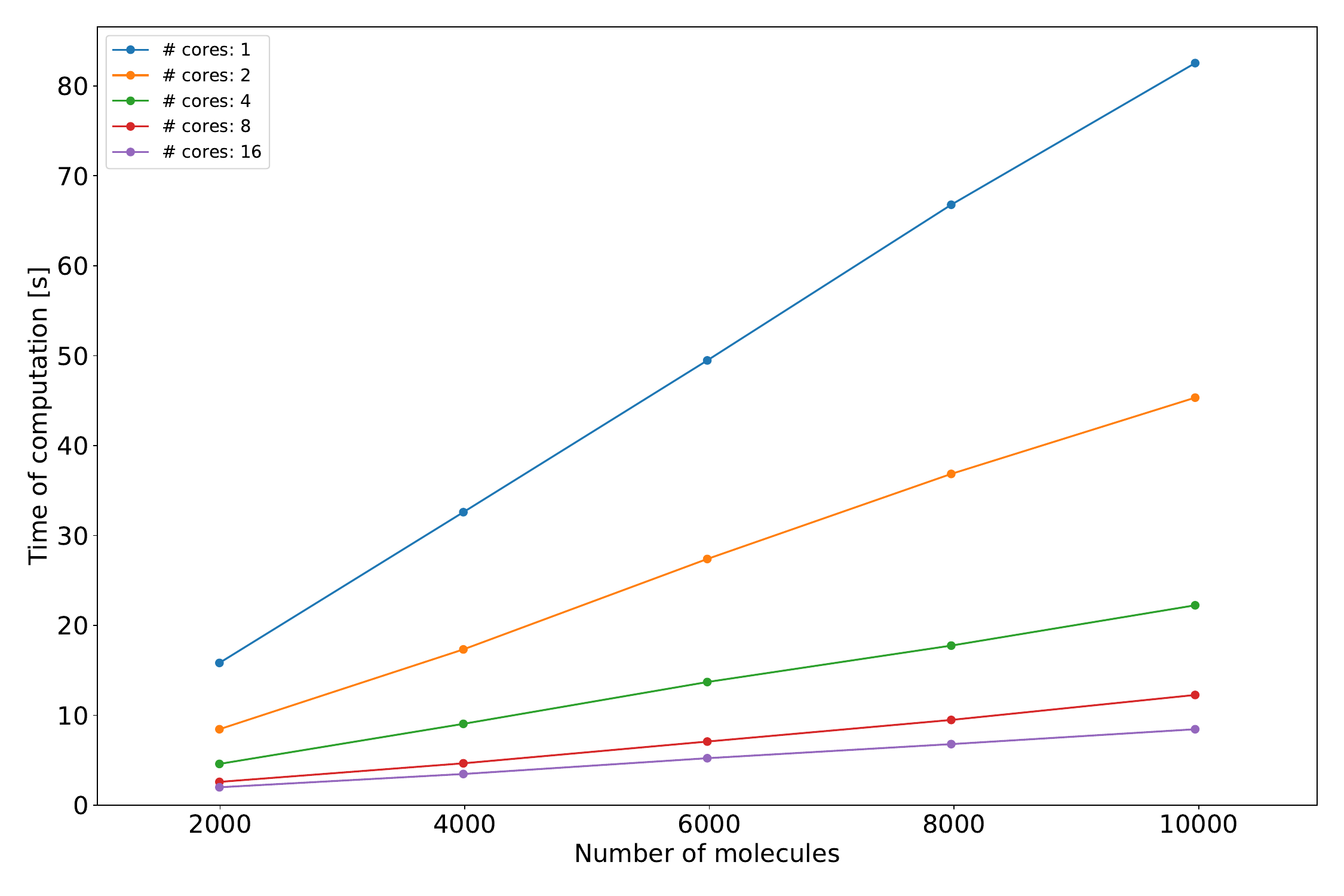}
    \caption{Computation time for PubChem fingerprint.}
    \label{fig:time-pubchem}
\end{figure}

\begin{figure}
    \centering
    \includegraphics[width=\textwidth]{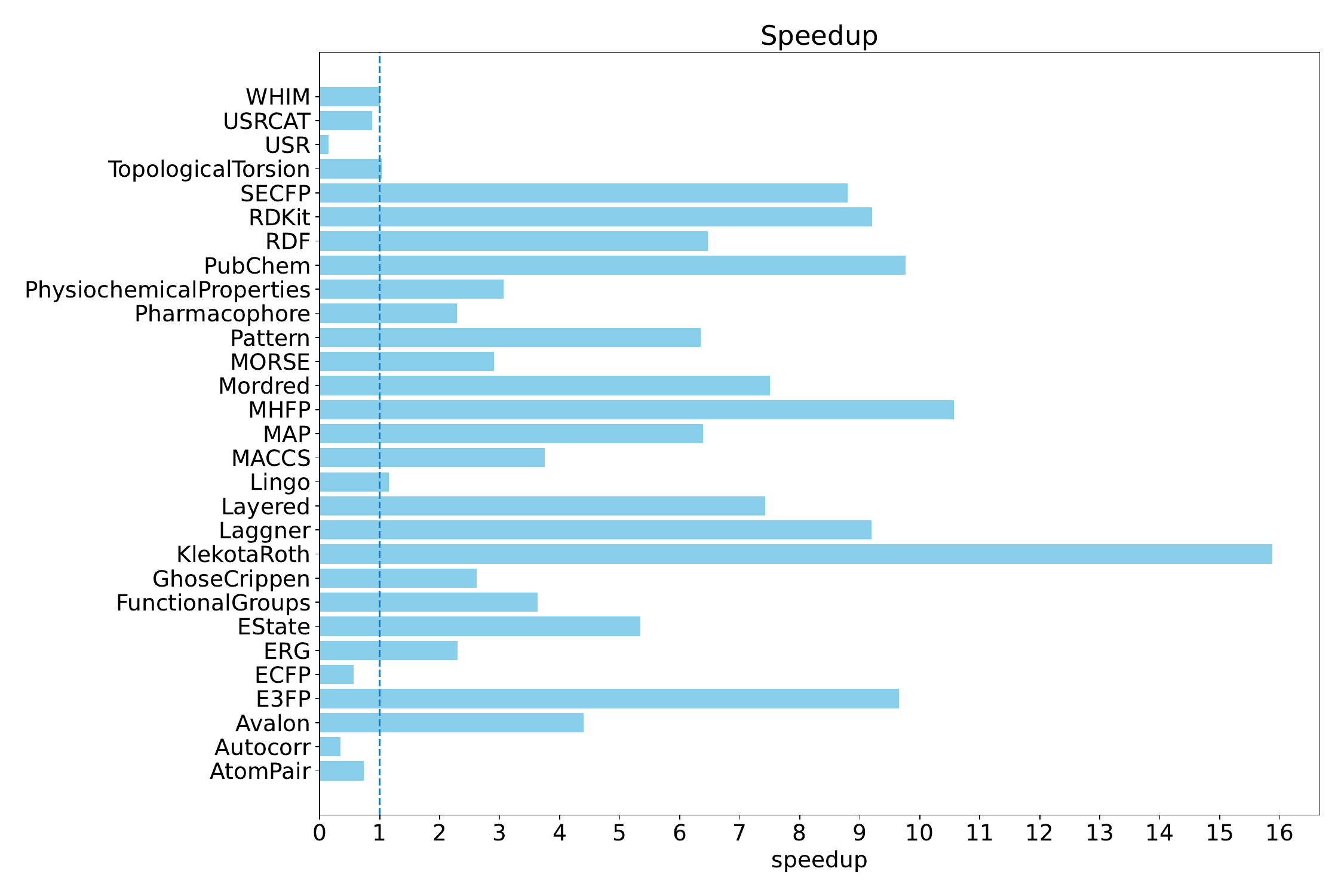}
    \caption{Speedup for fingerprints when using 16 cores.}
    \label{fig:fp_16_cores_speedups}
\end{figure}

As an example, we present the timings for the PubChem fingerprint \cite{fp_pubchem}, commonly used for virtual screening, in Figure \ref{fig:time-pubchem}. Speedup for all fingerprints \footnote{We omit Pharmacophore fingerprint due to excessive computation time. Due to the checking of multiple SMARTS patterns for all atoms, it is by far the slowest fingerprint.} is shown in Figure \ref{fig:fp_16_cores_speedups}, when using 16 cores and 10 thousand molecules. Speedup is defined as a ratio of sequential to parallel computation time. We calculate those times as an average of 5 runs, using a machine with Intel Core i7-13700K 3.4 GHz CPU. For 3D fingerprints, we do not include the conformer generation time.

PubChem fingerprint clearly benefits from parallelism, with time clearly decreasing when using more cores. This behavior is typical for more computationally heavy fingerprints, like substructure-based ones, which have to check numerous SMARTS patterns for each molecule. In particular, as visible in Figure \ref{fig:time-pubchem}, this gain appears for many data sizes. Even for just 2000 molecules, the time decreases from about 15 seconds to just about 2 seconds, which is much more convenient for interactive analyses and ad hoc queries, like searching for similar molecules.

High speedup values indicate that a significant majority of fingerprints benefit from parallelism, with Klekota-Roth achieving the greatest improvement. In general, computationally expensive ones like SECFP or Mordred gain the most. Only the fastest fingerprints, like ECFP or Atom Pair, have a speedup less than 1, meaning slower computation than the sequential one. However, we did not tune the number of cores, and using 6 or 8 could be enough for some fingerprints given this amount of data.

\begin{figure}
    \centering
    \includegraphics[width=0.85\textwidth]{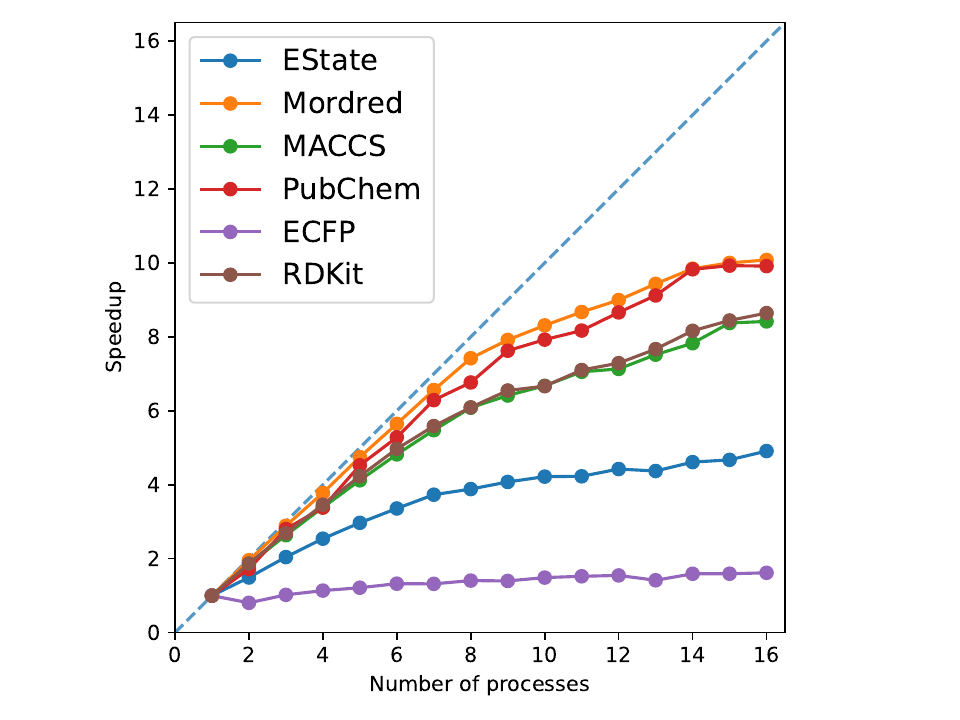}
    \caption{Speedup plot for selected fingerprints.}
    \label{fig:speedups}
\end{figure}

Lastly, in Figure \ref{fig:speedups} we provide a detailed speedup plot for six commonly used fingerprints of different types: hashed (ECFP \cite{fp_ecfp} and RDKit \cite{fp_rdkit}), substructural (MACCS \cite{fp_maccs} and PubChem \cite{fp_pubchem}), and descriptors (EState \cite{fp_estate} and Mordred \cite{fp_mordred}). Here, we could use the entire HIV dataset (about 41 thousand molecules), since the computational cost was much lower for only six fingerprints. Overall, fingerprints are well scalable with number of cores, particularly heavier ones like Mordred or PubChem. They achieve almost perfect linear speedup up to about 8 cores. Only the extremely fast ECFP fingerprint seems to be better suited for sequential computation, at least for a dataset of this size.

\subsection{Sparse matrix support}

Molecular fingerprints are often extremely sparse. Therefore, using proper representation can result in large memory savings, compared to dense arrays. Differences are particularly significant for large datasets, which are typical for virtual screening or similarity searching.

\skfp~has full support for sparse matrix computations, using SciPy. As an example, we calculated the memory usage of the resulting fingerprint arrays for PCBA dataset from MoleculeNet \cite{MoleculeNet}, consisting of almost 440 thousand molecules. In Table \ref{table:fingerprints-sparse-savings}, we report memory usage of dense and sparse representations. We also report memory savings, defined as how many times the sparse representation reduced the memory usage. For brevity, we show the results of 5 fingerprints with the largest reduction. Code to produce results for all fingerprints is available in the GitHub repository, in \texttt{benchmarking} directory.

Clearly, fingerprints greatly benefit from sparse representations, with a density of arrays around just 1-2\%. In particular, popular ECFP and FCFP fingerprints \cite{fp_ecfp} are among those that benefit the most. The Klekota-Roth fingerprint \cite{fp_klekota_roth}, which is quite long for a substructure-based fingerprint, obtains a reduction from almost 2 GB RAM to just 23 MB, i.e. 88.2 times. Those savings would be even more important during the hyperparameter tuning of downstream classifiers when many copies of the data matrix are created. Using a sparse representation did not negatively impact computation time, compared to the dense one.

\begin{table}[]
\centering
\resizebox{\textwidth}{!}{
\begin{tabular}{|c|c|c|c|}
\hline
\textbf{Fingerprint name} & \textbf{\begin{tabular}[c]{@{}c@{}}Dense array\\ size (MB)\end{tabular}} & \textbf{\begin{tabular}[c]{@{}c@{}}Sparse array\\ size (MB)\end{tabular}} & \textbf{Memory savings} \\ \hline
Klekota-Roth              & 2029                     & 23                        & 88.2x                     \\ \hline
FCFP                      & 855                      & 15                        & 57x                       \\ \hline
Physiochemical Properties & 855                      & 17                        & 50.3x                     \\ \hline
ECFP                      & 855                      & 19                        & 45x                       \\ \hline
Topological Torsion       & 855                      & 19                        & 45x                       \\ \hline
\end{tabular}
}
\caption{Memory usage of fingerprints in dense and sparse versions.}
\label{table:fingerprints-sparse-savings}
\end{table}

\subsection{Molecular property prediction}

\skfp~can greatly simplify the process of classifying molecules. We show a part of a pipeline in Listing \ref{code:molecular_property_prediction}, responsible for computing ECFP fingerprints from SMILES strings and their classification. For brevity, we omit loading the data, which is just standard Pandas code.

Inputs can be any sequences that consist of SMILES strings or RDKit \texttt{Mol} objects, e.g. Python lists or Pandas series. As \texttt{ECFPFingerprint} is a stateless transformer class, it uses an empty \texttt{.fit()} method in the pipeline. The code is also parallelized, requiring only the \texttt{n\_jobs} parameter.
\\

\lstset{basicstyle=\small}
\begin{lstlisting}[basicstyle=\ttfamily\footnotesize,language=Python, label=code:molecular_property_prediction]
from sklearn.ensemble import RandomForestClassifier
from sklearn.pipeline import make_pipeline

from skfp.fingerprints import ECFPFingerprint

pipeline = make_pipeline(
    ECFPFingerprint(n_jobs=-1),
    RandomForestClassifier(n_jobs=-1, random_state=0)
)
pipeline.fit(smiles_train, y_train)

y_pred = pipeline.predict(smiles_test)
\end{lstlisting}

\subsection{Fingerprint hyperparameter tuning}

Most papers in the literature neglect hyperparameter tuning for molecular fingerprints, only tuning downstream classifiers. We conjecture that this is also due to the lack of easy-to-use and efficient software for computing fingerprints. The works that perform such tuning \cite{fingerprints_hyperparameters,fingerprints_hyperparameters_2} indicate that it is indeed beneficial.

We perform hyperparameter tuning for all 2D fingerprints on MoleculeNet single-task classification datasets \cite{MoleculeNet}, using the scaffold split provided by OGB \cite{OGB}. Only the pharmacophore fingerprint was omitted due to the excessive computation time for some molecules. A Random Forest classifier with default hyperparameters was used, in order to isolate the tuning improvements to just fingerprints. In Table \ref{table:fingerprints-tuning}, we report the area under receiver operating characteristic curve (AUROC) values obtained when using tuned hyperparameters, improvement from tuning compared to the default parameters, and average gain over all datasets. Due to space limitations, we present the results for 5 fingerprints that had the highest average gain. They can therefore be considered as the methods with the highest tunability \cite{tunability}. The hyperparameter grids and code are available in the GitHub repository, in \texttt{benchmarking} directory.

Tuning fingerprints results in considerable gains, as high as 5.8\% AUROC in case of RDKit fingerprint \cite{fp_rdkit} on BBBP dataset. Notably, substructure-based Ghose-Crippen fingerprint \cite{fp_ghose_crippen} gains 4\% AUROC on average, using feature counts instead of binary indicators. This signifies that further research in this area, using \skfp, would be highly beneficial.

\begin{table}[t]
\centering
\resizebox{\textwidth}{!}{
\begin{tabular}{|c|ccc|c|}
\hline
\multirow{2}{*}{\textbf{Fingerprint}} & \multicolumn{3}{c|}{\textbf{Dataset AUROC and tuning gain}}                            & \multirow{2}{*}{\textbf{\begin{tabular}[c]{@{}c@{}}Average tuning\\ AUROC gain\end{tabular}}} \\ \cline{2-4}
                                      & \multicolumn{1}{c|}{\textbf{BACE}} & \multicolumn{1}{c|}{\textbf{BBBP}} & \textbf{HIV} &                                                                                               \\ \hline
GhoseCrippen                          & \multicolumn{1}{c|}{84.0 (+2.9)}   & \multicolumn{1}{c|}{73.3	(+4.9)}   & 76.0	(+4.3)  & +4.0                                                                                          \\ \hline
RDKit                                 & \multicolumn{1}{c|}{83.0 (+1.2)}   & \multicolumn{1}{c|}{73.0 (+5.8)}   & 76.7	(+0.6)  & +2.5                                                                                          \\ \hline
Laggner                               & \multicolumn{1}{c|}{80.1	(+3.1)}   & \multicolumn{1}{c|}{73.8	(+0.7)}   & 76.1 (+1.0)  & +1.6                                                                                          \\ \hline
Avalon                                & \multicolumn{1}{c|}{83.8	(+2.3)}   & \multicolumn{1}{c|}{71.3	(+0.6)}   & 78.0 (+1.7)  & +1.5                                                                                          \\ \hline
EState                                & \multicolumn{1}{c|}{82.3 (+1.7)}   & \multicolumn{1}{c|}{71.7 (+1.0)}   & 76.4 (+0.0)  & +0.9                                                                                          \\ \hline
\end{tabular}
}
\caption{Molecular property prediction performance using different fingerprints and gain from tuning their hyperparameters.}
\label{table:fingerprints-tuning}
\end{table}

\subsection{Complex pipelines for 3D fingerprints}

For tasks requiring 3D information, i.e. fingerprints based on conformers, the whole processing pipeline becomes more complex. Conformers need to be generated and often post-processed with force field optimization, and resulting fingerprints may have missing values. Additionally, using more than one fingerprint is often beneficial, especially for virtual screening, as they take into account different geometry features. In Listing \ref{code:3d_pipeline}, we present an example of how to create such a pipeline to vectorize molecules for screening, calculating the GETAWAY \cite{fp_getaway} and WHIM \cite{fp_whim} descriptors. This short example would require well over 100 lines of code in RDKit, even without parallelization.
\\

\begin{lstlisting}[basicstyle=\ttfamily\footnotesize,language=Python, label=code:3d_pipeline]
from sklearn.impute import SimpleImputer

from skfp.fingerprints import (
    GETAWAYFingerprint, WHIMFingerprint
)
from skfp.preprocessing import ConformerGenerator
from sklearn.pipeline import make_pipeline, make_union


pipeline = make_pipeline(
    ConformerGenerator(
        optimize_force_field="MMFF94", n_jobs=-1
    ),
    make_union(
        GETAWAYFingerprint(n_jobs=-1),
        WHIMFingerprint(n_jobs=-1)
    ),
    SimpleImputer(strategy="mean"),
)
\end{lstlisting}

\subsection{Comparison with existing software}

We compare \skfp~with existing libraries for chemoinformatics, which also support the computation of molecular fingerprints. Differences are summarized in Table \ref{table:cdk-openbabel-comparison}.

In terms of Python support, we provide the first Python-native solution, with other libraries relying on various wrappers. It is also installable with \texttt{pip} from PyPI, and can be easily managed with modern dependency managers such as Poetry \cite{Poetry}. We implement the largest number of fingerprints, including both all those available in other libraries, and new ones like MAP4 \cite{fp_map4} or E3FP \cite{fp_e3fp}. Another advantage of \skfp~is the full support of parallelism and even distributed computing, which is nonexistent or very limited elsewhere. It is also the only library utilizing pre-commit hooks, dedicated security tools, and offering a fully scikit-learn compatible interface.

\begin{table}[!ht]
\resizebox{\textwidth}{!}{
\begin{tabular}{|c|c|c|c|c|}
\hline
 & \textbf{CDK} & \textbf{Open Babel} & \textbf{RDKit} & \textbf{scikit-fingerprints} \\ \hline
Language & Java & C++ & C++ & Python \\ \hline
pip-installable & No & Yes & Yes & Yes \\ \hline
Last PyPI update & Never & 2020 & 2024 & 2024 \\ \hline
Number of fingerprints & 13 & 7 & 22 & 31 \\ \hline
scikit-learn compatible & No & No & No & Yes \\ \hline
Parallelism & No & No & Very limited & Yes \\ \hline
Pre-commit hooks & No & No & No & Yes \\ \hline
Code quality tools & Yes & No & Yes & Yes \\ \hline
Security tools & No & No & No & Yes \\ \hline
Integrated datasets & No & No & No & Yes \\ \hline
Easy proprietary usage & \begin{tabular}[c]{@{}c@{}}Yes\\ (LGPL-2.1)\end{tabular} & \begin{tabular}[c]{@{}c@{}}No\\ (GPL-2.0)\end{tabular} & \begin{tabular}[c]{@{}c@{}}Yes\\ (BSD-3)\end{tabular} & \begin{tabular}[c]{@{}c@{}}Yes\\ (MIT)\end{tabular} \\ \hline
\end{tabular}
}
\caption{Comparison of scikit-fingerprints with other solutions.}
\label{table:cdk-openbabel-comparison}
\end{table}

\section{Impact}

\skfp~is a comprehensive library for computing molecular fingerprints. Leveraging fully scikit-learn compatible interfaces, researchers can easily integrate it with complex pipelines for processing molecular data. Comprehensive capabilities, with over 30 fingerprints, both 2D and 3D, with efficient conformer generation, enable using varied solutions for molecular property prediction, virtual screening, and other tasks. Intuitive and unified APIs make it easy to use for domain specialists with less programming expertise, like computational chemists, chemoinformaticians, or molecular biologists. We also put strong emphasis on code quality, security, and automated checks and analyzers.

The lack of efficient parallelism is a major downside of existing solutions. Modern molecular databases can easily encompass millions of molecules, especially for virtual screening \cite{virtual_screening,virtual_screening_2}. Our solution, utilizing all available cores, results in significant speedups, enabling efficient processing of large datasets. This is also beneficial for hyperparameter tuning \cite{fingerprints_hyperparameters, fingerprints_hyperparameters_2}, fingerprint concatenation \cite{fingerprints_concatenation}, data fusion \cite{data_fusion,data_fusion_2}, and other computationally complex tasks.

Simple class hierarchy and high code quality make our solution easily extensible. New fingerprints can be easily added, automatically benefiting from parallelization and scikit-learn compatibility. GitHub repository had 7 contributors to date, showing a good reception by the community and an easy learning curve. The first issue by an external researcher has been made in a week of making the library public, highlighting the need for modern software in this area.

Research shows that fingerprint-based molecular property prediction remains competitive compared to graph neural networks \cite{fingerprints_applications,fingerprints_applications_2,fingerprints_applications_4}, justifying further research in this area. In particular, they should be applied as baselines for a fair evaluation of the impact of novel approaches, which is particularly easy with our library. \skfp~has already been applied to molecular chemistry research. In \cite{MOLTOP}, it was used to implement ECFP fingerprint as a baseline algorithm, ensuring fair comparison of various approaches on the MoleculeNet benchmark. It is also actively applied to predict the toxicity of pesticides for honey bees, using the recently proposed ApisTox dataset \cite{ApisTox}. Furthermore, numerous research projects at the Faculty of Computer Science at AGH University of Krakow are currently utilizing it.

Finally, \skfp~is constantly evolving, with new fingerprints being added. We are also working on expanding the functionality, e.g. implementing data splitting functions based on fingerprints, or adding molecular filters like Lipinski's Rule of 5 \cite{lipinski} for preprocessing. Therefore, its impact in chemoinformatics will be even greater in the future.

\section{Conclusions}

We have developed \skfp, an open source Python library for computing molecular fingerprints. It is simple to use, fully compatible with the scikit-learn API, and easily installable from PyPI. It is also the most feature-rich and highly efficient library available in the open source Python ecosystem, allowing parallel computation of more than 30 different fingerprints. Multiple mechanisms have been implemented to ensure high code quality, maintainability, and security. It fills the gap for a single, definitive software in the Python ecosystem for molecular fingerprints. It facilitates quicker, more efficient, and more comprehensive experiments in the fields of chemoinformatics, drug design, and computational molecular chemistry.

\section*{Acknowledgements}
\label{}

Research was supported by the funds assigned by Polish Ministry of Science and Higher Education to AGH University of Krakow, and by the grant from \q{Excellence Initiative - Research University} (IDUB) for the AGH University of Krakow. The authors thank Michał Szafarczyk and Michał Stefanik for their help with code implementation, Wojciech Czech for help with manuscript review, and Andrew Dalke for helpful discussions and comments. The authors also thank Aleksandra Elbakyan for her work and support for accessibility of science.

\bibliographystyle{elsarticle-num} 
\bibliography{bibliography}

\clearpage

\section*{Appendix: scikit-mol comparison}
\label{}

This is an additional section, added post-publication and not contained in the published journal version.

Here, we also compare to the scikit-mol library \cite{scikit-mol}, summarized in Table \ref{table:scikit-mol-comparison}.

\begin{table}[h!]
\centering
\begin{tabular}{|c|c|c|}
\hline
\textbf{} & \textbf{scikit-mol} & \textbf{scikit-fingerprints} \\ \hline
Language & Python & Python \\ \hline
pip-installable & Yes & Yes \\ \hline
Last PyPI update & 2023 & 2024 \\ \hline
Number of fingerprints & 8 & 31 \\ \hline
scikit-learn compatible & Yes & Yes \\ \hline
Parallelism & Yes & Yes \\ \hline
Pre-commit hooks & No & Yes \\ \hline
Code quality tools & No & Yes \\ \hline
Security tools & No & Yes \\ \hline
Integrated datasets & No & Yes \\ \hline
Easy proprietary usage & \begin{tabular}[c]{@{}c@{}}Yes\\ (LGPL-3.0)\end{tabular} & \begin{tabular}[c]{@{}c@{}}Yes\\ (MIT)\end{tabular} \\ \hline
\end{tabular}
\caption{Comparison of scikit-fingerprints with scikit-mol.}
\label{table:scikit-mol-comparison}
\end{table}

Both libraries are written in Python, are installable with \texttt{pip}, and offer a scikit-learn compatible API. In terms of licensing, both enable easy proprietary usage.

While both libraries parallelize fingerprints computation, scikit-mol relies on the built-in Python multiprocessing, which serializes all objects with Pickle when passing them between processes. In \skfp, we use the Joblib implementation, which optimizes the serialization and deserialization of NumPy arrays. They are the main result of fingerprint calculation, and the largest objects passed between processes. Hence, their optimization may result in efficiency gains, particularly for large datasets. Joblib also allows seamless switching to distributed computation, since it supports Dask clusters as computation backend.

\skfp~has much larger implementation scope, with almost four times the number of fingerprints in scikit-mol, and integration of commonly used datasets. Code quality tools, pre-commit hooks, and security tools are not used in scikit-mol, whereas \skfp~includes them in the CI/CD process.

\end{document}